\documentclass[amsmath,superscriptaddress,amssymb]{revtex4}

\begin{document}

\def\beqa{\begin{eqnarray}}
\def\eeqa{\end{eqnarray}}
\def\beqn{\begin{equation}}
\def\eeqn{\end{equation}}

\def\g{g}
\def\gb{{\bar \g}}
\def\wz{w}
\def\D{D}                 
\def\Gam{\Gamma}          
\def\Gxx{{\Gamma^-_{\s\s}}}   
\def\Gpxx{{\Gamma^+_{\s\s}}}
\def\Gyy{{\Gamma_{\sy\sy}}}
\def\Gst{[\Gamma]_{st}}
\def\Gan{\delta \Gamma}
\def\R{R}                 
\def\RE{E}
\def\Rb{{\bar \R}}
\def\REb{{\bar \RE}}
\def\REs{(\RE^0)}
\def\REt{(\RE^1)}
\def\W{W}
\def\Wb{{\bar \W}}
\def\PbN{\delta {\bar \Phi}_N}   
\def\PbP{\delta {\bar \Phi}_P}
\def\pst{\phi}
\def\V{V}        

\def\pro{\pi}
\def\pros{(\pro^0)}
\def\prot{(\pro^1)}

\def\x{x}                  
\def\xxm{\x_1^-}
\def\xxp{\x_1^+}
\def\xy{\x_2}
\def\xb{\mathbf{\x}}       
\def\xbx{\xb_1}
\def\xbxm{{\xb_1^-}}
\def\xbxp{{\xb_1^+}}
\def\xby{{\xb_2}}
\def\Partial{\nabla}       
\def\Deltb{{\bar \Delta}}  
\def\t{t}                  
\def\txm{\t_1^-}
\def\txp{\t_1^+}
\def\ty{\t_2}
\def\s{s}                  
\def\sxm{\s_1^-}
\def\sxp{\s_1^+}
\def\sx{\s_1}
\def\sy{\s_2}      
\def\u{u}               
\def\uy{\u_2}  
\def\uxm{\u_1^-}
\def\uxp{\u_1^+}
\def\a{a}                  
\def\aan{\delta \a}
\def\aant{\delta \a_\td}
\def\aang{\delta \a_\Gamma}

\def\u{u}                  
\def\ub{{\bar \u}}
\def\uan{\delta \u}
\def\T{T}                  
\def\r{r}                  
\def\rx{{\r_1}}
\def\ry{{\r_2}}
\def\lr{{\bar \r}}                 
\def\ang{\varphi}          
\def\angx{{\ang_1}}
\def\angy{{\ang_2}}
\def\Dang{\Delta \ang}      

\def\k{k}                  
\def\ks{\mathbf{\k}}
\def\n{n}                  
\def\nxm{\n_1^-}
\def\nxp{\n_1^+}
\def\nym{\n_2^-}
\def\nyp{\n_2^+}

\def\td{{\cal T}}          
\def\tdxy{{\cal T}_{12}}         
\def\tdan{\delta \td}            

\def\S{S}
\def\Sr{{\S_\r}}
\def\DSr{\Delta \Sr}
\def\St{s}
\def\ri{\rho}              
\def\rib{{\bar \ri}}

\def\ynu{y}
\def\ynuan{\delta \ynu}

\def\T{T}                   
\def\E{E}                  
\def\J{J}                  
\def\San{\delta \Sr}       
\def\eb{{e}}               
\def\ux{p}                
\def\db{{\sigma}}          
\def\Om{\Omega}

\def\f{f}                  
\def\intc{\oint}   
\def\I#1{I\lbrace{#1}\rbrace}
\def\Isst#1{I_\S\lbrace{#1}\rbrace}
\def\Ipst#1{I_\ang\lbrace{#1}\rbrace}

\def\M{M}                  
\def\m{m}                  
\def\Gauss{\kappa}
\def\GN{G_N}
\def\GP{G_P}
\def\cN{\zeta_N}
\def\cP{\zeta_P}
\def\Gts{{\tilde G}^0}
\def\Gtt{{\tilde G}^1}

\def\bE{\beta}
\def\cE{\gamma}
\def\dE{\delta}

\def\chE{(\chi)}
\def\dchE{(\delta \chi)}

\def\c{c}                  
\def\MS{{M_\circ}}         
\def\lrM{r_{M}}         
\def\vP{v_P}               

\def\stand#1{\left[#1\right]_\mathrm{st}}

\def\ln{\mathrm{ln}}       
\def\sect#1{sect.\,#1}       
\def\model{{\rm{model}}}

\title{Post-Einsteinian tests of gravitation}
\author{ Marc-Thierry  Jaekel}

\affiliation{Laboratoire de Physique Th\'eorique de l'ENS,
24 rue Lhomond, F75231 Paris Cedex 05 \footnote{
Centre National de la Recherche Scientifique (CNRS), Ecole Normale Sup\'{e}rieure
 (ENS), Universit\'{e} Pierre et
Marie Curie (UPMC); email:jaekel@lpt.ens.fr}}

\author{Serge Reynaud }

\affiliation{Laboratoire Kastler Brossel, case 74, Campus Jussieu,
F75252 Paris Cedex 05 \footnote{CNRS, ENS, UPMC;
email:reynaud@spectro.jussieu.fr}}


\begin{abstract}
Einstein gravitation theory can be extended by preserving its geometrical 
nature but changing the relation between curvature and energy-momentum tensors. 
This change accounts for radiative corrections, replacing the Newton 
gravitation constant by two running couplings which depend on scale 
and differ in the two sectors of traceless and traced tensors.
The metric and curvature tensors in the field of the Sun, which were 
obtained in previous papers within a linearized approximation, are then 
calculated without this restriction.
Modifications of gravitational effects on geodesics are then studied, 
allowing one to explore phenomenological consequences of extensions 
lying in the vicinity of general relativity. 
Some of these extended theories are able to account for the Pioneer anomaly 
while remaining compatible with tests involving the motion of planets.
The PPN Ansatz corresponds to peculiar extensions of general relativity 
which do not have the ability to meet this compatibility challenge. 


\end{abstract}
\maketitle

\section{Introduction}

Gravity teets which have been performed in the solar system show a fairly 
good agreement with general relativity (GR) \cite{Will}.
The geometrical basis of GR, that is essentially the equivalence principle, 
is one of the most accurately verified properties of the physical world.
The confrontation of observations with predictions of parametrized 
post-Newtonian extensions of GR shows that such an extension cannot
escape a small neighbourhood of GR.
Tests of the gravity force law also result in stringent bounds on potential 
deviations from the standard form of GR, from the millimiter range to the 
size of planetary orbits \cite{Fischbach}.
There are however a few indications which may be seen as challenging GR.

Observations at galactic and  cosmological scales have led to introduce 
large amounts of dark matter first, of dark energy then, in order to fit
the predictions of GR. As long as these dark components are not seen by
other means, the observations can as well be attributed to modifications 
of the gravitational force law at very large scales 
\cite{Perlmutter,Aguirre,Riess,Sanders02,Lue04,Turner04}.
This idea can be tested by following the motions of deep space probes 
reaching the outer part of the solar system and the first probes of this kind
have in fact shown anomalous accelerations \cite{Anderson98,Anderson02}. 
The anomaly has been observed as unexplained residuals of the Doppler tracking 
data from Pioneer 10/11 probes with respect to predictions of GR.
They are usually described as an anomalous acceleration directed towards the Sun 
with an approximately constant amplitude $\simeq 0.8\,{\rm{nm\,s^{-2}}}$ over a 
large range, 20 to 70 astronomical units (AU), of heliocentric distances.
Up to now, the anomaly has escaped all attempts of explanation as a systematic 
effect generated by the spacecraft or its environment \cite{Anderson03,Nieto04}.
This status has motivated a lot of studies aimed at a better understanding 
of deep space navigation as well as investigating the possibility that the 
Pioneer anomaly be the first hint of modified gravity laws at large scales 
\cite{Turyshev04,Bertolami04}, 
and led to proposals for new missions \cite{Pioneer05}.

A new kind of extensions of GR has recently been proposed with the remarkable 
property that they are able to account for the Pioneer anomaly while remaining 
compatible with other gravity tests performed in the solar system
\cite{JR05mpl,JR05cqg}. This feature has been analyzed in the context of a 
linearized treatment of gravitation fields, which restricts its applicability.
It is the aim of the present paper to overcome this restriction by 
studying the new framework without assuming such a linearization. 
Before going into the heart of the subject, it is worth recalling the motivations 
for a generalization of general relativity, which are rooted in its now long standing 
confrontation with Quantum Field Theory (see more references in \cite{JR05mpl,JR05cqg}).

When considered as a field theory, GR appears to be a good effective theory at 
intermediate scales where accurate observations or experiments are available.
However, radiative corrections due to the coupling of gravity field to the other
fields naturally lead to consider GR as immersed in a family of fourth order theories 
\cite{deWitt,Deser74,Capper74} with different behaviours at shorter or longer scales. 
This larger family shows the advantage of exhibiting renormalizability as well as 
asymptotical freedom at high energies \cite{tHooft,Stelle,Fradkin}. 
This suggests that extensions of GR can in principle be tackled by using the powerful 
tools of modern renormalization group equation \cite{Reuter02,Reuter04,Reuter05}. 
A complete solution of renormalization group trajectories appears as a
formidable task at the moment but preliminary studies already confirm that
GR can be a considered as the limit at intermediate ranges of a well defined
field theory with different behaviours at shorter and longer ranges.
As this research program is still far from completion, an alternative approach is 
to explore potential extensions of GR from a phenomenological point of view. 

This view meets the aims of the parametrized post-Newtonian (PPN) formalism 
\cite{Will}: theories lying in the vicinity of GR may indeed be explored through 
the anomalies they can produce in gravitational observations or experiments. 
Of course, these theories have to obey an important criterium of compatibility 
with the stringent bounds set by the variety of tests already performed in the 
solar system. As we shall see in the following, PPN models are peculiar
cases of the more general extensions studied in the present paper.
They cannot account for the Pioneer anomaly while remaining 
compatible with planetary tests \cite{Anderson02}, in contrast to
the general extensions of GR studied here \cite{JR05mpl,JR05cqg}.

These extensions can be characterized as `post-Einsteinian' for the two following 
reasons. First, the core of Einsteinian theory of gravity is fully preserved~: 
the gravitational field is identified with a metric tensor $\g_{\mu\nu}$ and 
motions are described by associated geodesics.
As a consequence of this geometrical foundation of the theory, 
the weak equivalence principle is unaffected. 
As is well known, this principle is well verified at distances ranging from the 
millimeter in laboratory experiments \cite{Adelberger03} to the sizes of 
Earth-Moon \cite{Williams96} or Sun-Mars orbits \cite{Hellings83,Anderson96}.
The relative accuracy, better than $10^{-12}$ for some of these tests, 
is good enough to discard any interpretation of the Pioneer anomaly from a 
violation of the equivalence principle \cite{Anderson02,JR04}.
This does not mean that the equivalence principle is an exact property of nature,
but only that its potential violations \cite{Damour,Overduin00} are smaller
than the deviations from GR studied in the present paper.

Then, the coupling between curvature and energy-momentum tensors differs 
from its standard form in GR \cite{EinsteinHilbert}.
More precisely, the Einstein tensor $\RE_{\mu\nu}\equiv\R_{\mu\nu}-\g_{\mu\nu}\R$ 
and the stress tensor $\T_{\mu\nu}$ are simply proportional to each other in GR 
whereas they are related through a more general linear response in the new
`post-Einsteinian' theory. 
In other words, the Newton gravitation constant $\GN$ which expresses the 
proportionality of $\RE_{\mu\nu}$ to $\T_{\mu\nu}$ in GR is now 
replaced by two running coupling constants, which depend on scale and differ in 
the sectors of traceless and traced tensors. 
Note that $\RE_{\mu\nu}$ and $\T_{\mu\nu}$ are transverse tensors with their
linear coupling preserving this transversality \cite{Jaekel95}. 
The consequences of this modification have been studied in \cite{JR05mpl,JR05cqg} 
within a linearized approximation, but they will be investigated in the present 
paper without such a restriction.
The gravitational effects on orbits and light propagation will be calculated 
with the non-linearity of the theory fully taken into account.

As a result of such a study, we will show that the `post-Einsteinian' extensions 
of GR show the ability to account for the Pioneer anomaly while
remaining compatible with the gravity tests performed on planetary orbits.
The phenomenological discussion will be presented in terms of two potentials
representing the two running coupling constants in the space-time domain as in 
\cite{JR05mpl,JR05cqg}.
The first potential, which generalizes the usual Newton gravitation potential,
will have to remain very close to its standard form in order to preserve the
good agreement between GR and gravity tests performed on planetary orbits, 
either by checking out the third Kepler law, or by looking at the precession
of the perihelion of orbits \cite{Talmadge88}. 
Meanwhile, the second potential will open additional freedom and offer the possibility 
to accomodate a Pioneer-like anomaly for probes with eccentric trajectories. 
It will also lead to other observable anomalies, in particular in time delay, 
deflection or Doppler tracking experiments on electromagnetic sources passing 
behind the Sun \cite{Shapiro99,Iess99,Bertotti03,Shapiro04}. 

In the next sections we will present the modified gravitation equations (\sect{2})
and their solutions in terms of metric fields (\sect{3}), using the fact 
that the deviation from GR must remain small in order
to preserve compatibilily with solar system tests of gravity.
We will express the small variation of the post-Einsteinian theory from GR
in different manners, in particular two potentials representing the modifications 
of the metric fields, two independent components of Einstein curvature tensor 
which may now differ from zero outside the gravity source, 
or two running gravitation constants exhibiting a scale dependence. 
The relations between these equivalent representations will be written down
explicitly.

We will then study some potentially observable consequences of this small 
variation of the post-Einsteinian theory from GR, looking at geodesics followed
by probe masses (\sect{4}) as well as electromagnetic ranging between Earth and
such probes (\sect{5}). The Pioneer-like anomaly, which has been predicted to exist 
for probes having eccentric motions in the outer solar system \cite{JR05mpl,JR05cqg},
will be confirmed here without the restrictions of a linearized approach.
Perturbations of bound geodesics will be evaluated (\sect{6}), and the absence of
observation of anomalies of perihelion precession with respect to GR will be 
translated in terms of constraints on the deviations from GR met in the solar system.
Observational constraints will finally be gathered (\sect{7}), leading to  
the conclusion that the theoretical arguments presented in the beginning 
of the paper and the observational ones discussed then, when taken together, 
appear to favor a simple extension of GR  
introduced in \cite{JR05mpl,JR05cqg} as accounting for the Pioneer-like anomaly. 

\section{Modified gravitation equations}

As discussed in the introduction, we use the high accuracy of tests of the  
principle of equivalence to consider only metric extensions of GR. 
Moreover, we focus our attention on a static and isotropic metric representing 
space-time around a punctual stationary source.
In the solar system, this assumption amounts to disregard the effects of rotation 
and non sphericity of the Sun which affect significantly inner motions 
\cite{Shapiro89,effetNordtvedt} but have a small influence on outer ones. 
This assumption notably simplifies the description since metric fields only 
depend on two functions of one variable, the radius coordinate. 
This dependence will be written below in different coordinate systems, in particular 
Schwartzschild or Eddington isotropic coordinates commonly used to represent the 
Schwartzschild solution \cite{Landau,MTW} or its PPN extensions 
\cite{Eddington,Robertson,Schiff66,Nordtvedt68,Will72,Nordtvedt72}.

In this section, we write down these equivalent representations of the metric fields, 
together with their relations to curvatures.
We then deduce the form of the equations replacing those of GR, introducing anomalous 
gravitational constants in the two sectors of traceless and traced tensors which have
the same phenomenological content as anomalous parts of the metric.
Note that these relations, obtained in the static isotropic case, constitute 
a first approximation of actual anomalies.
They will have to be checked out in future works by performing more complete
calculations taking into account perturbations of the static isotropic case, 
due to the structure and rotation of the Sun as well as to the presence of planets
(see as an example the detailed calculation of the Pioneer anomaly within the
context of GR in \cite{Anderson02}).

A static isotropic metric may be characterized by two functions, corresponding to time 
and radial components written in terms of isotropic coordinates \cite{Will} 
\beqa
\label{Eddington_isotropic_metric}
&&d\s^2 = \g_{00}(\r)\c^2 d\t^2 + \g_{\r\r}(\r)\left( d\r^2
+ \r^2 (d\theta^2 + {\rm{sin}}^2\theta d\varphi^2)\right)\nonumber\\
&&\partial_0 \g_{\mu\nu} = \partial_\theta \g_{\mu\nu} =
\partial_\varphi \g_{\mu\nu} = 0 
\eeqa
(coordinates $x\equiv\left(\c\t,\r\cos\theta\cos\varphi,\r\cos\theta\sin\varphi,\r\sin\theta\right)$
with $\c$ the velocity of light).
The conventions are the same as in \cite{Landau} with
some definitions and general properties gathered in \ref{app:bianchi}.
An equivalent representation of the same spacetime may be given in terms of
Schwartzschild coordinates \cite{MTWchap23} 
\beqa
\label{Schwartzschild_isotropic_metric}
&&d\s^2 = \gb_{00}(\lr)\c^2 d\t^2 + \gb_{\r\r}(\lr)d\lr^2 
- \lr^2 \left(d\theta^2 + {\rm{sin}}^2\theta d\varphi^2\right)\nonumber\\
&&\partial_0 \gb_{\mu\nu} = \partial_\theta \gb_{\mu\nu} =
\partial_\varphi \gb_{\mu\nu} = 0 
\eeqa
The metric tensors, $\g_{\mu\nu}$ for the first coordinate system and $\gb_{\mu\nu}$ 
for the second one, do not depend on the variables $\t,\theta,\varphi$ which are the same 
in both systems. The relation between radial coordinates $\r$ and $\lr$ depends on the metric
itself
\beqa
\label{Eddington_Schwartzschild}
{d\r \over \r} &=& \sqrt{-\gb_{\r\r}} {d\lr \over \lr} =
\left(1 + {\r\partial_\r \g_{\r\r}\over 2\g_{\r\r}}\right)^{-1} {d\lr \over \lr}
\eeqa
Equations (\ref{Eddington_Schwartzschild}) also provide the functional relation
between the two different representations 
$\left(\g_{00}, \g_{\r\r}\right)$ and $\left(\gb_{00}, \gb_{\r\r}\right)$
of the degrees of freedom of gravitational field.

General expressions of the curvature tensors are given in  
\ref{app:bianchi} as well as Bianchi identities between these expressions.
As the static isotropic metric from which they are calculated, these curvatures 
contain only two independent components.
These two components may be chosen for example as the scalar curvature $\R$ and 
the single independent component $\W$ contained in the Weyl tensor 
\beqa
\label{Weyl_amplitude}
\W &\equiv& {\W^{0\r}}_{0\r} = {\W^{\varphi\theta}}_{\varphi\theta} \, , \quad
{\W^{0\theta}}_{0\theta} = {\W^{0\varphi}}_{0\varphi}
= {\W^{\r\theta}}_{\r\theta} = {\W^{\r\varphi}}_{\r\varphi} = - {\W\over 2} 
\eeqa
These two quantities are derived from the metric components, for example in terms of
isotropic coordinates, 
\beqa
\R(\r) &=& -{1\over \g_{\r\r}} \left( \left( \partial_\r + {2\over\r} \right)
\partial_\r \left(2 \ln\vert\g_{\r\r}\vert + \ln\g_{00}\right) \right.
\nonumber\\
&&+\left. {1\over2}\left(\left(\partial_\r \ln\vert\g_{\r\r}\vert \right)^2
+\partial_\r\ln\vert\g_{\r\r}\vert \partial_\r \ln\g_{00}
+\left( \partial_\r \ln\g_{00} \right)^2 \right) \right)\nonumber\\
\W(\r) &=& {1\over 3\g_{\r\r}} \left(\partial_\r -\partial_\r\wz -{1\over\r}\right)
\partial_\r\wz, \qquad
\wz \equiv {1\over2}{\ln }\left\vert{\g_{\r\r}\over \g_{00}}\right\vert
\eeqa
The advantage of this representation is that the scalar curvature $\R$ and Weyl 
amplitude $\W$ behave as scalars under reparametrizations of the radial coordinate
\beqa
\R(\r)&=& \Rb(\lr) \, , \qquad \W(\r)= \Wb(\lr) 
\eeqa
One may use equivalently another representation, with the two components 
$\RE^0_0$ and $\RE^\r_\r$ of Einstein tensor as independent degrees of 
freedom for gravitational field,
\beqa
\label{Einstein_curvature}
\RE^0_0(\r) &=& {1\over \g_{\r\r}} \left(
\left(\partial_\r + {2\over \r} \right) \partial_\r \ln\vert\g_{\r\r}\vert 
+ {1\over 4} \left( \partial_\r \ln\vert\g_{\r\r}\vert \right)^2  \right) \nonumber\\
&=& \REb^0_0(\lr) = {1\over\lr^2} \partial_\lr\left( \lr(1+{1\over\gb_{\r\r}})\right)
\nonumber\\
\RE^\r_\r(\r) &=& {1\over \g_{\r\r}} \left(
{1\over \r} \left( \partial_\r \ln\vert\g_{\r\r}\vert + \partial_\r \ln\g_{00} \right) +
 {1\over 4}  \partial_\r \ln\vert\g_{\r\r}\vert
\left(\partial_\r \ln\vert\g_{\r\r}\vert + 2\partial_\r \ln\g_{00} \right)
 \right) \nonumber\\
&=& \REb^\r_\r(\lr) = {1\over\lr^2} \left(
1 + {1\over \gb_{\r\r}}(1+ {\lr\partial_\lr\gb_{00}\over\gb_{00}})\right)
\eeqa
Note that the remaining components of the Einstein tensor $\RE^\theta_\theta$ 
and $\RE^\varphi_\varphi$, or $\REb^\theta_\theta$ and $\REb^\varphi_\varphi$,
are equal and determined by Bianchi identity
\beqa
\label{Einstein_Bianchi}
\Partial_\mu\RE^\mu_\r &=& \partial_\r \RE^\r_\r 
+ {1\over 2} \left( \partial_\r\ln\g_{00} \right) \left( \RE^\r_\r - \RE^0_0 \right)  
+ \left( \partial_\r\ln\vert\r^2g_{\r\r}\vert \right) 
\left(\RE^\r_\r -\RE^\theta_\theta\right) = 0
\eeqa

We now write generalized gravitation equations as a linear response relation between 
the Einstein curvature tensor and the energy-momentum tensor \cite{JR05mpl,JR05cqg} 
\beqa
\label{linear_response_equations}
\RE^\mu_\nu(x) &\equiv& \int {d^4x'}\, \chE^{\mu\rho}_{\nu\lambda}(x,x') 
\T^\lambda_\rho(x')
\eeqa
General relativity corresponds to the simple Einstein-Hilbert relation \cite{EinsteinHilbert},
that is also to a local expression for the linear response function
($\GN$ is the Newton gravitation constant),
\beqa
\label{Einstein_equations}
&&\stand{\RE^\mu_\nu} \equiv {8\pi \GN\over \c^4} \T^\mu_\nu \\
&&\stand{\chE^{\mu\rho}_{\nu\lambda}(x,x')} \equiv 
{4\pi \GN\over \c^4}\left( \delta^\mu_\nu \delta^\rho_\lambda 
+ \delta^\mu_\lambda \delta^\rho_\nu \right) 
\delta^{(4)}(x-x') \nonumber
\eeqa
Now the linear response equation, arising for example from radiative corrections of
GR through the coupling of gravity to quantum fields \cite{Jaekel95}, should contain 
a non local part, equivalent to a scale dependence of the response,
\beqa
&&\RE^\mu_\nu \equiv \stand{\RE^\mu_\nu} + \delta\RE^\mu_\nu \,,\qquad
\delta\RE^\mu_\nu(x) \equiv \int {d^4x'}\, \dchE^{\mu\rho}_{\nu\lambda}(x,x') 
\T^\lambda_\rho(x')
\eeqa
The integral equation then takes a simple form when the energy-momentum tensor 
of the source is static and punctual ($\M$ is the mass localized at the origin) 
\beqa
\label{modified_Einstein_equations}
&&\T^\mu_\nu(\x) = \delta^\mu_0\delta^0_\nu \M\c^2\delta^{(3)} (\mathbf{x}) \nonumber\\
&&\delta\RE^\mu_\nu(x) \equiv \M\c^2\int {\c dt'}\, \dchE^{\mu 0}_{\nu 0}(x,x')
\,,\qquad x'\equiv (\c\t',\mathbf{0})
\eeqa

In preceding equations, $\stand{\RE}$ denotes the standard GR solution and $\delta\RE$ 
the deviation of the post-Einsteinian solution from the standard one.
At this point, we must emphasize that observations in the solar system are compatible with GR, 
with the notable exception of Pioneer anomaly, and that deviations from GR are bound to remain 
very small \cite{Will}. We may also note that theoretical reasons point to the same conclusion, 
namely to restrict the attention to extensions remaining close to GR. 
As a matter of fact, radiative corrections show a better behaviour with respect to renormalization 
when computed in the vicinity of a solution of GR rather than in the vicinity of Minkowski 
metric \cite{tHooft}. Fourth order theories surrounding GR also show asymptotic freedom at
high frequencies \cite{Fradkin} and preliminary studies of renormalization group trajectories,
which are restricted to a single sector up to now \cite{Reuter02,Reuter04,Reuter05}, suggest 
that the full theory remains close to GR over a large range of length scales,
precisely those intermediate scales which are tested in the solar system \cite{Fischbach}. 

For these reasons, we will only consider extensions lying in the vicinity of GR.
This entails that the anomalous terms $\delta\RE$ or $\dchE$, as well as the anomalies to
be observed in gravitational phenomena, are small and may be calculated as perturbations 
around GR. As the latter corresponds to Einstein curvatures vanishing everywhere except 
on gravity sources, that is the Sun in the solar system, this assumption allows one to 
simplify the Bianchi identities (see the \ref{app:bianchi}).
Further simplifications arise when using the fact that potentials or curvatures have 
relatively small amplitudes in the solar system (see the \ref{app:approx}).
For example, the transversality of Einstein tensor then takes its simple Minkowskian form 
and the tensor can be decomposed on the sectors of traceless and traced tensors 
as in the linearized approximation \cite{JR05mpl,JR05cqg}.
The generalized gravitation equations may then be written in terms of two gravitation 
constants which differ in the two sectors and depend on momentum or length scale.
Two running coupling constants $\Gts=\GN+\delta\Gts$ and $\Gtt=\GN+\delta\Gtt$ thus 
replace the unique Newton gravitation constant $\GN$, introducing a dependence on the 
conformal weight, different in the two sectors, as well as on the length scale. 
The information contained in the anomalous running coupling constants $\delta\Gts$ and 
$\delta\Gtt$ is equivalently encoded in the spatial variation of the anomalous Weyl and 
scalar curvatures $\delta\W$ and $\delta\R$ of a solution or, also, of its two anomalous 
components $\delta\RE^0_0$ and $\delta\RE^\r_\r$. 

These equivalent formulations of the extensions of GR will be used in the following. 
General relations between them, assuming only the idea of perturbation in the vicinity
of GR, are to be found in the next section. 
These expressions, obtained around GR, preserve their full non linear dependence 
in Newton gravitation constant $\GN$.
Simpler relations are obtained in the linearized regime, which corresponds to a weak
Newton potential or, equivalently, to a perturbation around Minkowski spacetime.
An explicit correspondence may be deduced in this case between the running coupling 
constants $\delta\Gts$ and $\delta\Gtt$ and anomalies in the solutions  
(see the \ref{app:approx}).
In particular, these anomalies can be written in terms of two potentials 
which were introduced in the linearized approach of \cite{JR05mpl,JR05cqg}
but can be used also in the non linear approach of the present paper
(see the next section).

\section{Post-Einsteinian metric}

In this section, we discuss the post-Einsteinian metrics which represent the spacetime
solution for extensions of gravitation theory in the vicinity of GR.
These extensions are best characterized by the anomalous Einstein curvatures 
$\delta\RE_\mu^\nu$ which may now differ from zero apart from the gravity source. 
We express metric tensors in both systems of Schwartzschild or isotropic coordinates. 
As emphasized in the Introduction, we fully account for the non linearity of the 
relations between curvature and metric tensors.

We begin with the solution of the non perturbed theory, that is to say GR,
\beqa
&&\stand{\RE_\mu^\nu} =  8\pi \Gauss \delta^0_\mu \delta_0^\nu \delta^{(3)}(\xb)
\quad, \quad
\Gauss \equiv {\GN \M \over \c^2}
\eeqa
The constant $\Gauss$ is related to the Schwartzschild radius and the metric can be written 
in terms of Schwartzschild or isotropic coordinates (eqs \ref{Schwartzschild_isotropic_metric}
or \ref{Eddington_isotropic_metric}) with 
\beqa
\label{metric_st}
&&\stand{\gb_{00}} = 1-2\Gauss\ub =  -{1\over\stand{\gb_{\r\r}}} \quad , \quad 
\ub\equiv{1\over\lr} \\
&&\stand{\g_{00}} = \left({1-\Gauss\u/2\over1+\Gauss\u/2}\right)^2 \quad , \quad 
\u\equiv{1\over\r} \quad,\quad
\stand{\g_{\r\r}} = - (1+\Gauss\u/2)^4 = -{\u^2\over\ub^2} \nonumber
\eeqa
Metric components have been defined in terms of a single Newton potential 
$\Gauss\u$ or $\Gauss\ub$, however expressed in terms of different radial 
coordinates $\r$ or $\lr$ in the two coordinate systems;
inverse radial coordinates $\u$ and $\ub$ have been introduced.
All curvatures but Weyl one vanish outside the source for this GR solution
\beqa
\label{Einstein_solution}
&&\stand{\R} = \stand{\RE^0_0} = \stand{\RE^\r_\r} = 0\quad,\quad
\stand{\W} =  {\Gauss\u^3\over  \left(1+\Gauss\u/2\right)^6}
\eeqa

Extensions of GR are well characterized by the fact that Einstein curvatures 
no longer vanish apart from the gravity source.
The anomalous non zero values $\delta\RE_\mu^\nu$ are entirely characterized 
by two independent components, for example $\delta\RE^0_0$ and $\delta\RE^\r_\r$, 
the other components being given by the Bianchi identity (\ref{Einstein_Bianchi}) 
which is now read
\beqa
\label{Einstein_Bianchi_pert}
\partial_\r \delta\RE^\r_\r 
+ {1\over 2} \stand{ \partial_\r\ln\g_{00} } 
\left( \delta\RE^\r_\r - \delta\RE^0_0 \right)  
+ \stand{ \partial_\r\ln\vert\r^2g_{\r\r}\vert } 
\left(\delta\RE^\r_\r -\delta\RE^\theta_\theta\right)
&=& 0
\eeqa
$\delta\RE^0_0$ and $\delta\RE^\r_\r$ can be seen as representing the
anomalous running coupling constants (see the previous section).
They may be written in both coordinate systems, taking into account the 
fact that inverse radial coordinates ($\ub$, $\u$) satisfy relation 
(\ref{metric_st}) in GR, but a modified relation in extended theory 
\beqa
{\delta\ub\over\ub}(\u) &=&  {\delta\g_{\r\r}\over 2\stand{\g_{\r\r}}}
\eeqa

Anomalous metric components and Einstein curvatures are equivalent representations
of the extended theories. Their relation may be rewritten, using Schwartzschild
coordinates, 
\beqa
\label{Schwartzschild_metric_solution}
&&\gb_{00} = \stand{\gb_{00}} + \delta \gb_{00}\quad, \quad \gb_{\r\r} = \stand{\gb_{\r\r}}
+ \delta \gb_{\r\r}\nonumber\\
&&{\delta\gb_{\r\r}\over\stand{\gb_{\r\r}}}= -{\ub\over\stand{\gb_{00}}}
\int{\delta\REb_0^0\over\ub^4} d\ub\nonumber\\ 
&&{\delta\gb_{00}\over\stand{\gb_{00}}} =\int {d\ub\over\stand{\gb_{00}}^2}
\int^\ub{\delta\REb_0^0\over\ub^{\prime4}} d\ub^\prime 
+ \int{\delta\REb^\r_\r\over \ub^3}{d\ub\over \stand{\gb_{00}}} 
\eeqa
For the forthcoming discussions, it will be convenient to introduce 
two potentials $\PbN$ and $\PbP$ through the definitions
\beqa
\label{Schwartzschild_solution}
&&\delta\gb_{\r\r} = {2\ub\over(1-2\Gauss\ub)^2}(\PbN-\PbP)^{\prime} \quad, \quad
\f^\prime \equiv \partial_\ub\f \nonumber\\
&&\delta\gb_{00} = 2(1-2\Gauss\ub)
\int{\PbN^\prime -2\Gauss\ub\PbP^\prime\over(1-2\Gauss\ub)^2}d\ub 
\eeqa
Let us also note the following  relation, which will appear useful in the following,
\beqa
\label{identity}
\delta\gb_{00}^\prime &=& 2\PbN^\prime +2\Gauss\delta(\gb_{00}\gb_{\r\r})
\eeqa
In the linearized approximation, all corrections in $\Gauss\ub$ may be disregarded
and the simple relations used in \cite{JR05mpl,JR05cqg} are recovered.
Clearly the full equations (\ref{Schwartzschild_solution}) describe 
non linear effects of the Newton potential and have therefore a larger domain 
of validity than their linearized expression. 
The precise form of the non linear version (\ref{Schwartzschild_solution}) has
been chosen so that they lead to simple expressions for the anomalous
Einstein curvatures 
\beqa
\label{curvature_solution}
\delta\REb^0_0 &\equiv& 2\ub^4(\PbN -\PbP)^{\prime\prime}\nonumber\\ 
\delta\REb^\r_\r &\equiv& 2\ub^3\PbP^\prime
\eeqa

Finally similar relations may be written in terms of isotropic coordinates
\beqa
\label{Eddington_solution}
{\delta\g_{\r\r}\over\stand{\g_{\r\r}}} &=& -2(1-2\Gauss\ub)^{1\over2}
\int{(\PbN-\PbP)^\prime\over(1-2\Gauss\ub)^{3\over2}}d\ub\nonumber\\
{\delta\g_{00}\over\stand{\g_{00}}} &=& 2\int {\PbN^\prime- 2\Gauss\ub\PbP^\prime
\over(1-2\Gauss\ub)^2}d\ub -{2\Gauss\ub\over(1-2\Gauss\ub)^{1\over2}}
\int{(\PbN-\PbP)^\prime\over(1-2\Gauss\ub)^{3\over2}}d\ub
\eeqa
or, alternatively,
\beqa
\label{Eddington_metric_solution2}
{\delta\g_{\r\r}\over\stand{\g_{\r\r}}} &=& -(1-2\Gauss\ub)^{1\over2}
\int{d\ub\over(1-2\Gauss\ub)^{3\over2}}
\int^\ub{\delta\REb^0_0\over\ub^{\prime 4}}d\ub^\prime\nonumber\\
{\delta\g_{00}\over\stand{\g_{00}}} &=& \int {d\ub\over(1-2\Gauss\ub)^2}
\int^\ub{\delta\REb_0^0\over\ub^{\prime4}} d\ub^\prime
+ \int{\delta\REb^\r_\r\over \ub^3}{d\ub\over (1-2\Gauss\ub)}\nonumber\\
&&-{\Gauss\ub\over(1-2\Gauss\ub)^{1\over2}}
\int{d\ub\over(1-2\Gauss\ub)^{3\over2}}
\int^\ub{\delta\REb_0^0\over\ub^{\prime4}} d\ub^\prime
\eeqa

To sum up the results of this section, relations (\ref{Schwartzschild_solution}) 
and (\ref{Schwartzschild_metric_solution}) describe the
looked for correspondance between different representations of extensions of GR:
anomalous Einstein curvatures and Schwartzschild metric 
are given in terms of potentials $\PbN$ and $\PbP$ by relations which
take into account the non linear effects of standard Newton potential.
This feature will be preserved in the following sections,
when consequences will be derived for geodesic motions and light propagation. 
At this stage, it is worth comparing generalizations obtained in this manner with the 
widely used PPN Ansatz \cite{Eddington,Robertson,Schiff66,Nordtvedt68,Will72,Nordtvedt72}.
This PPN Ansatz is written in \ref{app:ppn} as a particular case of the more general 
extension presented here.
Non zero values are found for Einstein curvature $\delta\RE^\r_\r$ apart from the source,
at first order in $\Gauss$, but not for $\delta\RE^0_0$ which vanishes at the same order.
It follows that the PPN Ansatz does not have the ability to account for 
the Pioneer anomaly, since the latter is associated to a curvature $\delta\RE^0_0$ which
does not vanish. Moreover, the scaling laws are different for the PPN Ansatz and for
the extension accounting for the Pioneer anomaly (see \sect{7}).

\section{Modified geodesics }

We derive in this section the geometric elements associated with post-Einsteinian metric
which will be needed for discussing gravity tests.
As in the two previous sections, we consider extensions lying in the vicinity of GR
so that any expression will be obtained as the sum of a standard GR value
and of a small anomaly, to be treated as a first order perturbation around GR. 

As in any metric theory, motions obey the geodesic equation 
\beqa
\label{geodetic}
&&{\D u^\mu \over d\s} \equiv {d u^\mu \over d\s} 
+ \Gamma^\mu_{\nu\rho} u^\nu u^\rho = 0 
\eeqa
The covariant derivative $\D$ is determined by the Christoffel symbols $\Gamma^\mu_{\nu\rho}$ 
associated with the metric (\ref{Eddington_isotropic_metric}) or 
(\ref{Schwartzschild_isotropic_metric}) (see the \ref{app:bianchi} for definitions).
Accelerations undergone along the geodesics may be defined 
as Christoffel symbols projected along the motion
\beqa
\label{Christoffel}
\g_{\mu\lambda}{d\u^\lambda \over d\s} &=& 
 -\Gamma_{\mu,\nu\rho} \u^\nu \u^\rho \equiv  -\Gamma_{\mu,\s\s}
\eeqa
Each geodesic is contained in a constant plane which may be chosen
as $\theta = {\pi\over 2}$. A given geodesic is then determined by two 
constants of motion, the energy $\E$ and angular momentum $\J$, for a test mass $\m$
\beqa
\label{motion_constants}
\m\c^3\g_{00} {d\t \over d\s} &=& \E \quad , \quad
\m\c \g_{\r\r} \r^2{d\ang \over d\s} = \J 
\eeqa
Accelerations are then written, using isotropic coordinates,
\beqa
\label{acceleration_field}
\Gamma_{\r,\s\s} 
&=&-{1\over2\g_{00}}\left( \partial_\r\g_{00}-\partial_\r(\g_{00}\g_{\r\r})
\left({d\r\over d\s}\right)^2 -\partial_\r\left({\g_{00}\over\r^2\g_{\r\r}}\right) 
{\J^2\over\m^2\c^2}\right)\nonumber\\
\Gamma_{\ang,\s\s} 
&=&\left({2\over\r}+ {\partial_\r\g_{\r\r}\over\g_{\r\r}}\right)
{\J\over\m\c}{d\r\over d\s}
\eeqa 
They may also be expressed as sums of a standard GR expression 
and an anomaly, written at first order in the deviation from GR,  
\beqa
\label{anomalous_acceleration_field}
 \Gamma_{\r,\s\s} &=& \stand{\Gamma_{\r,\s\s}} + \delta \Gamma_{\r,\s\s}, \qquad
\Gamma_{\ang,\s\s} = \stand{\Gamma_{\ang,\s\s}} + \delta \Gamma_{\ang,\s\s}\nonumber\\
\delta \Gamma_{\r,\s\s} &=&
 -{1\over2}\partial_\r \left({\delta\g_{00}\over\stand{\g_{00}}}\right) \nonumber\\
&&+{1\over2}\left(\partial_\r\left({\delta\g_{00}\over\stand{\g_{00}}}\right)\stand{\g_{\r\r}}
+{\partial_\r\stand{\g_{00}}\over\stand{\g_{00}}}\delta\g_{\r\r} 
+ \partial_\r\delta\g_{\r\r} \right)\left({d\r\over d\s}\right)^2 \nonumber\\
&&+\left(\partial_\r\left({\delta\g_{00}\over\stand{\g_{00}}} -
{\delta\g_{\r\r}\over\stand{\g_{\r\r}}}\right) 
- {\r^2\over\stand{\g_{00}}} 
\partial_\r\left({\stand{\g_{00}}\over\r^2\stand{\g_{\r\r}}}\right)
\delta\g_{\r\r}\right)
{\J^2\over2\m^2\c^2\stand{\g_{\r\r}}\r^2}\nonumber\\
\delta \Gamma_{\ang,\s\s} &=& \partial_\r\left({\delta\g_{\r\r}\over\stand{\g_{\r\r}}}\right)
{\J\over\m\c}{d\r\over d\s}
\eeqa
Note that inserting the explicit form of metric perturbations 
(\ref{Eddington_metric_solution2}) leads to expressions of perturbations of geodesics 
in terms of the anomalous Einstein curvatures.

In order to discuss electromagnetic ranging, we will also need to express the
perturbations of time delays.
The latter are represented by a two-point function $\td$ describing the elapsed time along
the propapagation of a lightlike signal from a spatial point $(\rx,\angx)$ to another one 
$(\ry, \angy)$. Equivalently, this function represents the propagation in a static 
isotropic metric of a null geodesics taking place in the plane $\theta = {\pi\over 2}$.
This problem has been solved in its full generality in \cite{JR05cqg}
and we reproduce here the solution in terms of integrals along null geodesics
\beqa
\label{null_geodesic_time_delay}
&&\angx - \angy = \int_\rx^{\ry} {\rib d \r/\r^2 \over
\sqrt{e^{2\wz(\r)} - {\rib^2\over\r^2}}} \quad,\quad
e^{2\wz(\r)} \equiv -{\g_{rr}\over\g_{00}}(\r) \nonumber\\
&&\tdxy \left( \rx,\angx;\ry,\angy \right) = \int_\rx^\ry {e^{2\wz(\r)} 
d\r \over \c \sqrt{e^{2\wz(\r)} - {\rib^2\over\r^2}}}
\eeqa
The two endpoints of the null geodesic have been supposed to lie
on the same side of the gravity source. When this is not the case, the time delay can
be written as the sum of the delays between each endpoint and the point of closest 
approach. In all cases, the latter is determined in a self-consistent manner 
from the two endpoints by (\ref{null_geodesic_time_delay}).
Note that the integrals (\ref{null_geodesic_time_delay}) only depend on the
ratio $\left(-\g_{rr}/\g_{00}\right)$ as a consequence of conformal invariance
of null geodesics propagation.

Expression (\ref{null_geodesic_time_delay}) of the time delay function may then be expanded 
as the sum of the standard GR expression and of an anomaly calculated at first order
\beqa
\label{perturbed_time_delay}
&&c\tdxy = \c\stand{\tdxy} + \c\delta \tdxy \quad, \quad
\c\delta \tdxy =  \int_\rx^\ry \delta \wz (\r) \c d\stand{\td}\nonumber\\
&& \c d\stand{\td} \equiv { \stand{e^{2\wz(\r)}} d\r \over \sqrt{ \stand{e^{2\wz(\r)}} 
- {\rib^2\over\r^2}}} \quad , \quad
\delta \wz = {1\over2}\left( {\delta \g_{\r\r}\over \stand{\g_{\r\r}}}
- {\delta \g_{00}\over \stand{\g_{00}}}\right)
\eeqa
$\c \stand{\td}$ is the standard optical distance in GR and $\c \delta\td$
the first order variation in the extended theory. This deviation only 
depends on the perturbation $\delta \wz$ of the conformally invariant ratio 
$\left(-\g_{rr}/\g_{00}\right)$. The relation (\ref{perturbed_time_delay})
translates into quantitative terms the qualitative idea of a dephasing of
phasefronts by the anomalous Weyl curvature.

\section{Pioneer-like anomaly}

We come now to an essential property of the new framework, namely the prediction of 
a Pioneer-like anomaly for probes following eccentric trajectories in the outer 
solar system.
For that purpose, we first recall the results obtained in \cite{JR05cqg} 
for writing the effects of the modified metric on Doppler shifts.
We do not repeat the part of calculations which already included 
non linear effects of Newton potential.

Pioneer probes were followed through a Doppler velocity built up on
radio signals exchanged with a station on Earth: 
an up-link radio signal emitted by the station at $\xxm = (\c\txm, \xbxm)$
and received by the probe at $\xy= (\c\ty,\xby)$ was instantaneously transponded to
a down-link radio signal then received by the station at $\xxp=(\c\txp,\xbxp)$.
Both Earth and probe followed geodesics, with respective velocities $\u_1^\pm, \uy$.
The gauge invariant Doppler observable, denoted $\ynu$, is defined as the ratio of 
clock rates of a reference clock at the departure and arrival of the radio signal 
\cite{JR05cqg}
\beqa
\label{frequency_shifts}
e^{\ynu} &\equiv& {d \sxp \over d\sxm} \quad,\quad ds^2\equiv\g_{\mu\nu} d\x^\mu d\x^\nu
\quad,\quad u^\mu\equiv{d\x^\mu \over ds}
\eeqa
The Pioneer effect is observed when looking at the Doppler shift (\ref{frequency_shifts}) 
as a function of time. As in \cite{Anderson02}, we write it as an acceleration 
\beqn
\label{shift_derivative}
2 \a d\s \equiv -\c^2 d \ynu
\eeqn
The definition of $\a$ depends on the choice of the reference clock delivering $d\s$ whereas 
the definition of $\a d\s$ does not depend on this choice. 

All these observables may be determined from the time delay $\tdxy$ on up- and down-links
and on the motions of the endpoints.
Motions of Earth and probe have vanishing covariant accelerations $\D\uxm = \D\uxp = \D\uy = 0$, 
corrections associated with the position of stations on Earth being taken into account separately
(these small corrections are supposed to be linearly superposed to the anomaly and 
to disappear in the end of the calculations).
With these assumptions, the following exact relation is obtained \cite{JR05cqg}
\beqa
\label{Doppler_acceleration}
{2\a d\s \over \c^2} &=& -{\left(\uxm \cdot \D\nxm \right) \over(\uxm \cdot\nxm)} +
{\left(\uxp \cdot \D\nxp \right) \over(\uxp \cdot\nxp)} 
-{\left(\uy \cdot \D\nyp \right) \over(\uy \cdot\nyp)} +
{\left(\uy \cdot \D\nym \right) \over(\uy\cdot \nym)} 
\eeqa
$\n^\pm_a$ represent the directions of wavevectors at endpoints 
and $\cdot$ denotes a 4-dimensional scalar product
\beqa
\label{directions_waves}
&&(\n^\pm_1)_0 = - (\n^\pm_2)_0 = 1 \quad , \quad
(\n^\pm_a)_i = \mp \c {\partial \tdxy\over \partial (\xb_a^\pm)^i} \nonumber\\
&&(\u \cdot \n) \equiv \u^\mu \n_\mu
\eeqa
 
The Doppler acceleration (\ref{Doppler_acceleration}) has a twofold dependence 
versus the metric. A first dependence is due to the Christoffel symbols which enter 
the covariant derivatives and a second one to the variation of the time delays 
which enter expressions (\ref{directions_waves}). When evaluating the Pioneer anomaly,
that is to say the variation of $a$ due to the metric perturbation with respect to GR,
we thus obtain the anomaly as a sum of two terms
\beqa
\aan &=& \aant  + \aang 
\eeqa
$\aant$ corresponds to the variation of the time delays on radio links whereas
$\aang$ enters through the variation of the probe trajectory \cite{JR05cqg}
\beqa
\label{extra_acceleration}
{2\aant d\sy\over\c^2} &=& d \lbrace {1\over (\uy \cdot {\nym})}{\c d\tdan(\xbxm,\xby) 
\over d\sy}+
 {1\over (\uy \cdot {\nyp})}{\c d\tdan(\xbxp,\xby) \over d\sy}\rbrace
\nonumber\\
{2\aang d\sy\over\c^2} &=& d\sxm {(\delta \Gxx\cdot\nxm)\over(\uxm \cdot\nxm)}
- d\sxp {(\delta \Gpxx \cdot\nxp) \over (\uxp\cdot \nxp)} \\
&+& d\sy \left( {(\delta \Gyy \cdot\nyp) \over (\uy \cdot\nyp)}
- {(\delta \Gyy \cdot\nym) \over (\uy \cdot\nym)} \right) \nonumber
\eeqa
We have specified the reference clock to be on the probe $d\s=d\sy$, 
$d\sx^\pm$ being associated with the clock on the station orbit. 
Calculating at first order in the deviation from GR, we have replaced all 
quantities but $\tdan$ and $\delta\Gamma$ by their zeroth order approximation.

From now on, we furthermore specify that the Pioneer probes lie and move in the radial 
direction.
Using (\ref{acceleration_field}), one sees that the contributions of the
Christoffel symbols for the Earth may be neglected.
The anomalous accelerations (\ref{extra_acceleration}) may thus be rewritten
\beqa
\label{acceleration_anomaly_approximation}
{2\aant \over \c^2} &\simeq&   {d^2\over d \sy^2}\left(\c d\tdan(\xbxm,\xby)
+ \c d\tdan(\xbxp,\xby)\right) \nonumber\\
{2\aang \over \c^2} &\simeq&  -2\delta \Gamma_{\ry,\sy \sy}
\eeqa
Inserting explicit expressions of time delays (\ref{perturbed_time_delay})
and Christoffel symbols (\ref{anomalous_acceleration_field}),
we rewrite these two contributions in terms of metric perturbations
(functions are evaluated at the probe radial coordinate $\ry$)
\beqa
\label{anomalous_acceleration}
{2\aant \over \c^2} &\simeq& -{d\over d\sy}\left(\left({\delta\g_{00}\over\stand{\g_{00}}} 
-{ \delta\g_{\r\r}\over\stand{\g_{\r\r}}}\right)\sqrt{-{\stand{\g_{\r\r}}\over\stand{\g_{00}}}}
{d\ry\over d\sy}\right) \nonumber\\
{2\aang \over \c^2} &\simeq& \partial_\r \left({\delta\g_{00}\over\stand{\g_{00}}}\right) \\
&-&\left(\partial_\r\left({\delta\g_{00}\over\stand{\g_{00}}}\right)\stand{\g_{\r\r}}
+{\partial_\r\stand{\g_{00}}\over\stand{\g_{00}}}\delta\g_{\r\r}
+ \partial_\r\delta\g_{\r\r} \right)\left({d\ry\over d\sy}\right)^2 \nonumber 
\eeqa

Using expressions (\ref{Eddington_solution}), we may finally obtain
the anomalous accelerations in terms of the two potentials $\PbN$ and $\PbP$.
This leads to the non linear generalization of the expression obtained 
for the Pioneer anomaly in the linearized approximation \cite{JR05cqg}.
This expression is recovered in the particular case of probes located 
in the outer solar system, where Newton potential is small.
We will use it in the end of this paper for discussing the significance
of observational constraints on the post-Einsteinian extensions.

\section{Perihelion precession}

We now study the influence of post-Einsteinian extensions of GR 
on the bound geodesics followed by planets.
We discuss in particular the perihelion precession, an observable
well known in tests of gravitation theory \cite{Will}.
As for other observables studied in this paper, we focus our attention
on anomalies induced by variations of the metric around the standard GR situation.

As the non linearity of gravitation theory plays a key role
in the problem of perihelion precession, we present quite detailed
calculations, starting from the Hamilton-Jacobi equation written, for a static 
isotropic metric, in terms of time $\t$, azimuth angle $\ang$ and a phase shift $\Sr$
($\E$ and $\J$ being the constants of motion (\ref{motion_constants}))
\beqa
\label{Hamilton_Jacobi}
&&\S = -\E \t + \J \ang + \Sr\nonumber\\
&&(\partial_\r \Sr)^2 = -\g_{\r\r}\left({\E^2\over\c^2\g_{00}}-\m^2\c^2\right)
-{\J^2\over\r^2}
\eeqa
To analyse bound orbits, it is convenient to use Schwartzschild coordinates and rewrite 
the quantities (\ref{Hamilton_Jacobi}) in terms of corresponding metric components
\beqa
\label{massive_orbit}
\Sr &=& \mp \J \int^{1\over\lr}\sqrt{-{\gb_{\r\r}\over\gb_{00}}}
\sqrt{{\E^2\over\J^2\c^2} -\gb_{00}\left({\m^2\c^2\over\J^2} +\ub^2\right)}{d\ub\over\ub^2}
\\
\ang &=& \pm  \int^{1\over\lr} {\sqrt{-\gb_{00}\gb_{\r\r}}  \over
\sqrt{{\E^2\over\J^2\c^2} - \gb_{00}({\m^2\c^2\over\J^2} + \ub^2)}}d\ub
\quad,\quad \t = {\E \over\J\c^2} \int^\ang {d\ang\over\gb_{00}\ub^2} \nonumber
\eeqa
We recall that $\ub$ is the inverse of the radial coordinate $\lr$ and that the 
prime symbol used below denotes derivation with respect to $\ub$.

One obtains in particular the orbital period, that is to say the duration between 
two successive identical values of $\varphi$, as 
\beqa
\label{period}
\T &=& {\E \over\J\c^2}\int_0^{2\pi}{d\ang\over\gb_{00}\ub^2}
\eeqa
For circular orbits with a constant radial velocity, $\t$ and $\ang$ are proportional to each 
other, and the orbital period only depends on the metric component $\gb_{00}$. 
Equations (\ref{massive_orbit}) and (\ref{period}) thus lead to the third Kepler law
($\ub$ being the inverse radius of the circular orbit)
\beqa
&&\T = {2\pi\over\c} \left({-2\over\gb_{00}^\prime}\right)^{1\over2}\ub^{-{3\over2}}
\quad,\quad
{\m^2\c^2\over\J^2} +\ub^2 = {\E^2\over\J^2\c^2} {1\over \gb_{00}}
= -{2\ub\over \gb_{00}^\prime}\gb_{00}
\eeqa
Hence, the orbital period may be written as the sum of a standard GR expression and of
a post-Einsteinian anomaly 
\beqa
\label{Kepler3}
&&\T = \stand{\T} + \delta\T \quad, \quad
\stand{\T} = 2\pi \Gauss^{-{1\over2}}\ub^{-{3\over2}} \quad, \quad
{\delta \T\over \stand{\T}} =  -{\delta\gb_{00}^\prime \over 2\stand{\gb_{00}}^\prime}
\eeqa
These relations entail that third Kepler law depends on variations of $\gb_{00}$ only. 
In other words, bounds resulting from third Kepler law constrain a particular linear
combination of the two independent possible variations.
We will come back on this discussion later.

We now study perihelion precessions which can be determined from the variation $\Dang$ 
of the azimuth angle between two identical values of the radial coordinate $\lr$,
the elapsed time being nearly but not exactly one orbital period.
This quantity may thus be obtained from the phase shift $\Sr$ (\ref{massive_orbit})
evaluated over this elapsed time
\beqa
\label{action_precession}
\DSr &\equiv& \mp \J \intc
\sqrt{-{\gb_{\r\r}\over\gb_{00}}}
\sqrt{{\E^2\over\J^2\c^2} -\gb_{00}({\m^2\c^2\over\J^2} +\ub^2)}{d\ub\over\ub^2}
\nonumber\\
\partial_\J \DSr &=& 2\pi + \Dang
\eeqa
We now evaluate the variation of these quantities for a perturbation around GR, the 
constants of motion $\E, \J$ being kept fixed. We obtain the following anomalies for 
the phase shift and precession angle, at first order in the metric perturbation,
\beqa
\label{perturbed_Einstein_geodesic}
\DSr &=& \stand{\DSr} + \delta\DSr \quad,\quad
\Dang = \stand{\Dang} + \delta\Dang\nonumber\\
\delta\DSr &=& -{1\over2} \intc\stand{\gb_{00}} \delta \gb_{\r\r}d\stand{\Sr}
+ {\E^2\over2\J\c^2}\intc{\delta\gb_{00}\over\stand{\gb_{00}}^2\ub^2}d\stand{\ang}
\nonumber\\
\delta\Dang &=& \partial_\J \delta\DSr
\eeqa

For the nearly circular orbits followed by planets, the anomalous precession
$\delta\Dang$ may be expanded in the orbit eccentricity $\eb$.
This expansion is presented in \ref{App:precession} and we give here its result
for the post-Einsteinian anomaly of the perihelion precession
\beqa
{\delta\Dang\over\pi} &\simeq& -\left( \delta\gb_{\r\r}
-{\ub\delta\gb_{00}^{\prime\prime}\over\stand{\gb_{00}}^\prime}
+{\eb^2\ub^2\over4}\left(\delta\gb_{\r\r}^{\prime\prime}
-{\ub\delta\gb_{00}^{(4)}\over2\stand{\gb_{00}}^\prime}\right)\right)
\eeqa
Inserting explicit expressions for metric perturbations 
(\ref{Schwartzschild_solution},\ref{identity})
in terms of the potentials $\PbN$ and $\PbP$ (with $\PbN$ considered to be
 much smaller than $\PbP$), we also obtain
\beqa
\label{perihelion_constraint}
{\delta\Dang\over 2\pi} &\simeq& \ub\left(\ub\PbP\right)^{\prime\prime} 
- {\ub\PbN^{\prime\prime} \over2\Gauss} 
+ {\eb^2\ub^2\over8}\left(\left(\ub^2\PbP^{\prime\prime}+\ub\PbP^\prime\right)^{\prime\prime} 
- {\ub\PbN^{(4)} \over2\Gauss}\right)
\eeqa
At the limit of nearly circular orbits $\eb\to 0$, the precession anomaly 
(\ref{perihelion_constraint}) takes a simple form depending only on
the second derivatives of the functions $\PbN$ and  $\ub\PbP$
\beqa
\label{perihelion_constraint_circular}
\left({\delta\Dang\over 2\pi}\right)_{\eb\to 0} &=& 
\left(2\Gauss\ub\PbP-\PbN\right)^{\prime\prime} {\ub\over 2\Gauss} 
\eeqa

We may recover the well known expression of the anomalous perihelion precession 
in the PPN formalism \cite{Will} by inserting in (\ref{perihelion_constraint_circular}) 
the particular expressions (\ref{PPN_potentials}) of $\PbN$ and $\PbP$ for the PPN metric.
Keeping only the leading term, we obtain the anomalous perihelion precession in
terms of anomalous parts of Eddington parameters $\beta$ and $\gamma$
\beqa
\label{PPN_precession}
\left({\delta\Dang\over2\pi}\right)^{PPN} &\simeq& \left(2(\gamma-1)-(\beta - 1)\right) 
\Gauss\ub
\eeqa
The more general expression (\ref{perihelion_constraint_circular}) may thus be thought of
as stating that the functions $\PbN^{\prime\prime}/2\Gauss^2$ 
and $(\ub\PbP)^{\prime\prime}/2\Gauss$ promote the anomalous PPN parameters $\bE-1$ and 
$\cE-1$ from the status of constants to that of space dependent functions. 
However this identification can only be a first rough description of the more 
general expressions obtained throughout the paper.
In the next section, we use these more general expressions to discuss how 
planetary tests fix bounds on potential post-Einsteinian extensions of GR.

\section{Observational constraints}

We now use the expressions obtained in previous sections to draw constraints on 
post-Einsteinian extensions of GR from observations or experiments performed in
the solar system. Any potential anomaly exhibits a particular dependence on the 
two functions of the radius coordinate used to parametrize the extensions. 
This freedom will be the key point allowing us to accomodate the Pioneer anomaly
besides the other gravity tests performed in the solar system.
The two independent functions can be chosen as the anomalous metric components 
$\left(\delta\g_{00},\delta\g_{\r\r}\right)$, the potentials $\left(\PbN,\PbP\right)$
or the Einstein curvatures $\left(\delta\RE_0^0,\delta\RE_\r^\r\right)$.
Relations between these representations have been given in \sect{3}, so that they
can be used equivalently for the calculations.
For discussing the significance of the tests, we will emphasize the parametrizaton 
in terms of anomalous Einstein curvatures since they describe in the clearest 
manner the deviation from GR.

We first discuss the Pioneer-like anomalous acceleration (\ref{anomalous_acceleration})
calculated for probes with an eccentric motion in the outer solar system.
At leading order, this expression can be linearized since non linear corrections
would involve a small Newton potential at large heliocentric distances.
We obtained in this manner a simplified expression \cite{JR05cqg}
\beqa
\label{linearized_anomalous_acceleration}
{\aan \over\c^2} &\simeq& {1\over2}\partial_\r\delta\g_{00}
-  \partial_\r\delta\g_{\r\r}\left({d\r\over d\s}\right)^2
- {1\over2}\left(\delta\g_{00} +\delta\g_{\r\r}\right){d^2\r\over d\s^2}
\nonumber\\
&\simeq& \partial_\r \PbN - 2 \partial_\r \left(\PbN -\PbP\right)\left({d\r\over d\s}\right)^2
-\left(2\PbN -\PbP\right){d^2\r\over d\s^2}
\eeqa
Now, the first contribution, which depends only on $\PbN$, can simply be interpreted as
resulting from a long range modification of the Newton force law, and it cannot explain the Pioneer
anomaly without being noticed on planetary tests \cite{Anderson02,JR04}.
Then, the last term proportional to the acceleration of the probes can be ignored, since the latter 
keep practically constant velocities over the observation range.
Finally, the term $\PbN$ can be neglected in front of $\PbP$ in the parenthesis appearing in 
the second term \cite{JR05mpl,JR05cqg}. We thus obtain
\beqa
&&\aan \simeq 2\vP^2 \partial_\r \PbP \quad,\quad
\vP \simeq 1.2 \times10^4 {\rm{m s}}^{-1}
\eeqa
where $\vP$ is the velocity of Pioneer probes. If the result is identified with the anomalous
acceleration recorded on Pioneer probes, we get an information about the function $\PbP$
at large distances 
\beqa
\label{Pioneer_constant}
&&\aan \simeq -8\times 10^{-10}{\rm{m s}}^{-2}
\quad,\quad
\c^2 \ub^2 \partial_\ub \PbP = -\c^2 \partial_\r \PbP \simeq 0.25 {\rm{m s}}^{-2}
\eeqa

Pioneer data \cite{Anderson02} indicate that the anomalous acceleration remains practically 
constant over a large distance range $20-70$ AU of heliocentric distances 
in the outer part of the solar system.
This suggests that the potential $\PbP$ has a linear dependence in the radial coordinate 
$\lr$ or, equivalently, varies as the inverse of $\ub$ over the same distance range.
This behaviour appears to favor a simple post-Einsteinian model \cite{JR05mpl,JR05cqg}
characterized by a single additional parameter $\cP$ with respect to GR, 
the latter being fixed by (\ref{Pioneer_constant}), 
\beqa
\label{Pioneer_model}
\left(\PbP\right)_\model \simeq -{\cP M\over\ub\c^2 }\quad,\quad
\cP M\simeq 0.25 {\rm{m s}}^{-2}&&
\eeqa
It should be emphasized that a similar component introduced in the 
potential $\PbN$ is constrained by planetary tests to remain vanishingly small
\beqa
\label{Newton_model}
\left(\PbN\right)_\model \simeq -{\cN M\over\ub\c^2 }\quad,\quad
\left\vert\cN M\right\vert \leq 10^{-12} {\rm{m s}}^{-2}&&
\eeqa
This estimation given in \cite{JR05cqg} should be refined after the discussions of the
present paper but the difference in the orders of magnitude is large enough
to ensure that $\PbN$ is negligible in front of $\PbP$ in the outer solar system.
We may then derive from (\ref{curvature_solution}) the expressions of the anomalous 
Einstein curvatures 
\beqa
\label{Einstein_model}
\left(\delta\RE^0_0 \right)_\model \simeq {4\cP M\ub\over\c^2 } \quad,\quad
\left(\delta\RE^\r_\r \right)_\model \simeq {2\cP M\ub\over\c^2 } &&
\eeqa

As already emphasized, this is the clearest manner to characterize the post-Einsteinian 
extension of GR studied in the present paper. 
In contrast to GR, the extended theory corresponds to Einstein curvatures with non null
values far from the gravity sources.
This feature is not a complete novelty since the PPN metric also shows Einstein curvatures
differing from their standard null values in free space surrounding the gravity sources.
But PPN expressions (\ref{PPN_curvatures}) exhibit peculiarities which entail
that they cannot account for the Pioneer anomaly.
First, they lead to a non null value of $\delta\RE^\r_\r$ at first order in $\Gauss$
but to $\delta\RE^0_0$ vanishing at the same order: $\delta\RE^0_0$ 
is negligible with respect to $\delta\RE^\r_\r$ in the PPN metric whereas the two 
quantities have comparable magnitudes in (\ref{Einstein_model}).
In other words, the PPN Ansatz explores a particular linear combination of the
two possible extensions which is not the linear combination producing
a Pioneer-like anomaly. Then, the anomalous Einstein curvatures obtained
in the PPN metric scale as $\ub^3$ for the leading term, which is again clearly 
distinct from those associated with the model (\ref{Pioneer_model}) which scale as $\ub$. 
In other words, the PPN metric which is defined as an expansion of the metric in terms 
of Newton potential can only produce short-range effects whereas the Pioneer anomaly
is rather a long-range modification of GR. Note that such a long-range effect can
only be obtained with an infrared modification of the running couplings 
\cite{JR05mpl,JR05cqg}.

The preceding discussions refer to the values of the two potentials $\PbN$ and $\PbP$ 
at large heliocentric distances, typically in the range $20-70$ AU where the Pioneer 
anomaly has been recorded. But the presence of the same potentials might also be
detected in the inner part of the solar system where very accurate measurements
have been performed. This question was discussed in the end of \cite{JR05cqg}
for the case of deflection experiments which can be analyzed within a linearized
framework. Here, we consider planetary tests bearing on
third Kepler law and perihelion precession \cite{Talmadge88}.
As already discussed, tests of the third Kepler law (\ref{Kepler3}) depend on 
variations of the temporal metric component $\gb_{00}$ only. 
At leading order, these variations amount to a modification of the Newton potential 
$\PbN$ which must remain very small in order to reproduce the agreement of GR
with planetary measurements \cite{Anderson02,JR04}. 
This leads to the bound (\ref{Newton_model}) written above 
for the function $\PbN$. 

The discussion for the bounds drawn from perihelion precessions involves not only the 
modification of the Newton potential $\PbN$ but also the second potential $\PbP$.
In fact the constraints on $\PbN$ prevent it to significantly contribute to the perihelion 
precession anomaly, so that (\ref{perihelion_constraint}) can be translated into
\beqa
\label{perihelion_estimations}
{\delta\Dang \over 2\pi}
\simeq \ub\left(\ub\PbP\right)^{\prime\prime}
+ {\eb^2\over8}\left(\ub^2\left(\ub(\ub\PbP)^{\prime\prime\prime}\right)^\prime 
+ \ub\left(\ub\PbP\right)^{\prime\prime} -2\ub\PbP^\prime\right)&&
\eeqa
We have kept the leading term which subsists at the limit of a circular orbit
as well as the sub-leading terms which depend on the squared eccentricity. 
Noticing that the dominant term in (\ref{perihelion_estimations}) vanishes 
for the simple model (\ref{Pioneer_model}) suggested by the Pioneer anomaly,
we deduce that the sub-leading one thus determines the perihelion precession
anomaly. To give an idea of the orders of magnitude, we write 
(\ref{perihelion_estimations}) for the simple model (\ref{Pioneer_model}), 
considering the Mars orbit (radius $\lr_M \sim 1.5$AU, eccentricity 
$\eb_M \sim .09$) and assuming that the parameter $\cP$ has the value
(\ref{Pioneer_model}) needed to fit the Pioneer anomaly
\beqa
\label{perihelion_model}
{\delta\Dang_M \over 2\pi} \simeq -{\cP M\lr_M\over\c^2 }{\eb_M^2\over4} 
\sim 1.2 \times 10^{-9}&&
\eeqa
This value is of comparable order of magnitude than the accuracy on Mars perihelion 
precession (see the bound deduced in \cite{JR04} from the data of \cite{Coy03}). 
This means that perihelion precession anomalies might be used to detect the effect
of the second potential in a dedicated analysis starting from the new model  
(\ref{Pioneer_model}). 

This analysis would have to look for values of the new parameters added with
respect to a PPN model: the first of these parameters is $\cP$
but the variation of $\cP$ versus distance should also be taken into account.
To make this point clear, let us consider a model differing from 
(\ref{Pioneer_model}) through the addition of other power law terms
\beqa
\left(\PbP\right)_\model &=& -\sum_{n} {\frac{\cP^{(n)}M\lr^{1+n}}{c^{2}}} \nonumber\\
\left(-c^{2}\partial _{r}\PbP\right)_\model  &=& \sum_{n} {(1+n)\cP^{(n)}M\lr^n}
\eeqa
The term $n=0$ in this sum reproduces (\ref{Pioneer_model}) and leads to a perihelion
precession anomaly proportional to the squared eccentricity. The other terms $n\neq 0$
lead to relatively larger contributions to the perihelion precession anomaly
\beqa
\left({\frac{\delta \Delta \varphi }{2\pi }}\right)_\model  = -\sum_n {\frac{\left( 1+n\right) {%
\zeta _{P}^{(n)}}M\lr^{n+1}}{c^{2}}}\left( n+{\frac{{e}^{2}}{8}}\left(
n+1\right) ^{2}\left( n+2\right) \right)  &&
\eeqa
To fix ideas, let us calculate the effect for the Mars orbit, taking into
account the next term $n=1$ besides the term $n=0$,
\beqa
\left({\delta\Dang_M \over 2\pi} \right)_\model \simeq -{\cP^{(0)} M\lr_M\over\c^2 }
\left( 2\epsilon + {\eb_M^2\over4} \right)\quad,\quad
\epsilon\equiv{\cP^{(1)}\lr_M\over\cP^{(0)}}&&
\eeqa
As already alluded to, the perihelion precession anomaly may thus be appreciably affected
(for values of $\epsilon$ of the order of ${\eb_M^2\over4}$) though the potential 
remains close to (\ref{Pioneer_model}) (because $\epsilon\ll 1$). 
In other words, the perihelion precession anomaly could be used as a sensitive
probe of the parameter $\cP$ and of its relative variation $\epsilon$. 
We stress once again that these values are evaluated at the radius of Mars orbit,
if the perihelion precession of Mars is studied, and not at the heliocentric
distances of Pioneer probes. Dedicated planetary tests could therefore bring to
our knowledge new informations on the potential deviations from GR, the aim
being to cover the intermediate distance range between the close vicinity of
the Sun, probed in deflection experiments \cite{JR05cqg}, and the outer solar
system, probed by Pioneer spacecrafts. 

\section{Conclusion}

In this paper, we have presented a new kind of extensions of GR
which allows one to explore potential variations of gravitational 
theory.
These extensions, which are motivated by the radiative corrections 
arising from the coupling of gravity field to the other fields, 
have been explored from a phenomenological point of view. 
Their ability to account for the Pioneer anomaly while remaining 
compatible with other gravity tests performed in the solar system,
which had been proved within a linearized treatment of gravitation 
fields \cite{JR05mpl,JR05cqg}, has been confirmed here without this 
restriction.

The new framework generalizes the PPN Ansatz \cite{Will} as well as the 
idea of a long-range modification of the gravity force law \cite{Fischbach}.
The new framework is characterized by two potentials which are functions
of the radius: the first one can be identified with a modification of
the gravity force law while the second one replaces the PPN constant $\cE$.
In fact, the first potential $\PbN$ can only slightly deviate from its standard
form, as a consequence of the good agreement of planetary tests with GR.
The key ingredient allowing one to account for the Pioneer anomaly is 
the second potential $\PbP$ with a non standard long range dependence, 
approximately a linear variation of the second potential versus the radius.
We have shown that the PPN metric tensor was a particular extension 
of GR which did not show the ability to account for the Pioneer anomaly,
in contrast to the general post-Einsteinian metric. 

The second potential $\PbP$, needed to reproduce a Pioneer-like anomaly, also 
leads to other observable anomalies, in particular in time delay, deflection 
or Doppler tracking experiments on electromagnetic sources occulted by the Sun.
Such anomalies have been analyzed in \cite{JR05cqg} and the results of this 
analysis are unchanged because a linearized treatment is sufficient to this aim. 
We may recall here that crucial indications could be obtained through a 
reanalysis of existing data, in particular those of the Cassini experiment 
\cite{Bertotti03}. On a longer term, they can be drawn from high accuracy 
Eddington tests (see for example the LATOR project \cite{Lator}) or global 
mapping of deflection over the sky (see for example the GAIA project \cite{Gaia}). 

In the present paper, we have studied the effect of the second potential $\PbP$
on the anomalous perihelion precession of planets, in particular Mars.
We have shown that dedicated analysis of planetary data could also bring
new information of interest on the variation of $\PbP$ at planetary ranges.
Another possible test of the new framework could be to confront its detailed 
predictions with the sets of Doppler tracking data recorded during the flight of 
Pioneer 10/11 probes towards the frontiers of the solar system. Such a
study might provide interesting clues, in particular through an inspection of
the variation of the anomaly during and after the flyby sequences which brought 
the probes to their escape trajectories.
Another attractive idea would be to confront the predictions of the new framework 
with the results of a specifically designed mission  \cite{Dsgp}. 

\appendix
\section{Vicinity of Einstein theory}
\label{app:bianchi}

In this appendix, we recall the general definitions of curvatures, with the conventions
of \cite{Landau}, and we write Bianchi identities obeyed by these curvatures.
We then deduce simplified relations valid for gravity theories lying in the vicinity
of GR.

We first reproduce the definitions of Christoffel symbols 
$\Gam^\lambda_{\mu\nu}$, Riemann ($\R_{\lambda\mu\nu\rho}$), Ricci ($\R_{\mu\nu}$)
and scalar ($\R$) curvatures
\beqa
\label{def_R_curvatures}
\Gam^\lambda_{\mu\nu} &\equiv& {\g^{\lambda\rho}\over2}
\left(\partial_\mu \g_{\nu\rho} +\partial_\nu \g_{\mu\rho} 
-\partial_\rho \g_{\mu\nu}\right)
\nonumber\\
\R^\lambda_{\mu\nu\rho} &\equiv& \partial_\nu \Gam^\lambda_{\mu\rho}
-  \partial_\rho \Gam^\lambda_{\mu\nu} +
\Gam^\lambda_{\nu\sigma} \Gam^\sigma_{\mu\rho} -
\Gam^\lambda_{\rho\sigma} \Gam^\sigma_{\mu\nu}\nonumber\\
\R_{\mu\nu} &\equiv& \Rb^\lambda_{\mu\lambda\nu}, \qquad 
\R\equiv\R^\mu_\mu
\eeqa
We also recall the definition of Einstein curvatures $\RE_{\mu\nu}$ 
\beqa
\label{def_E_curvatures}
\RE_{\mu\nu} &\equiv&\R_{\mu\nu} -{1\over2}\R, \qquad \RE\equiv -\R
\eeqa
and Weyl curvatures $\W_{\lambda\mu\nu\rho}$
\beqa
\label{def_W_curvatures}
\W_{\lambda\mu\nu\rho} &\equiv& \R_{\lambda\mu\nu\rho}
- {1\over2}\left(\g_{\lambda\nu}\R_{\mu\rho} + \g_{\mu\rho}\R_{\lambda\nu}
- \g_{\lambda\rho}\R_{\mu\nu} - \g_{\mu\nu}\R_{\lambda\rho}\right)
\nonumber\\&&
+{1\over6}\left(\g_{\lambda\nu}\g_{\mu\rho} -\g_{\lambda\rho}\g_{\mu\nu}\right)\R
\eeqa

Riemann curvatures express the commutators of translations of vector fields in a 
curved space \cite{MTWchap15}. 
They satisfy Bianchi identities, that is also Jacobi identities for the corresponding algebra,
\beqa
\label{Bianchi_R}
&&\Partial_\sigma \R^\lambda_{\mu\rho\nu} + \Partial_\nu \R^\lambda_{\mu\sigma\rho}
+ \Partial_\rho \R^\lambda_{\mu\nu\sigma} =0\nonumber\\
&&\Partial_\lambda \R^\lambda_{\mu\rho\nu} = \Partial_\rho \R_{\mu\nu} -
\Partial_\nu \R_{\mu\rho}\, ,\qquad
\Partial_\lambda \R^\lambda_{\mu} = {1\over2}\Partial_\mu \R
\eeqa
These identities entail in particular that Einstein curvature tensor is transverse
($\Partial_\mu$ is the covariant derivative associated with the metric)
\beqa
\label{Bianchi_E}
\Partial_\lambda \RE^\lambda_{\mu} &\equiv&
\partial_\lambda \RE^\lambda_{\mu} + \Gam^\lambda_{\lambda\rho}\RE^\rho_\mu -
\Gam^\rho_{\lambda\mu}\RE^\lambda_\rho=0 
\eeqa
They also lead to relations between covariant derivatives of the Weyl curvatures
and Einstein tensor 
\beqa
\label{Bianchi_W}
\Partial_\lambda \W^\lambda_{\mu\nu\rho} &=&
{1\over2}\Partial_\nu\left(\RE_{\mu\rho} -{1\over3}\g_{\mu\rho}\RE\right)
-{1\over2}\Partial_\rho\left(\RE_{\mu\nu} -{1\over3}\g_{\mu\nu}\RE\right)\nonumber\\
\Partial^\nu\Partial_\lambda \W^{\lambda}_{\mu\nu\rho} &=&
{1\over2}\Partial^2\RE_{\mu\rho} 
+{1\over6}\left(\Partial_\mu\Partial_\rho-\g_{\mu\rho}\Partial^2\right)\RE
- {1\over2}\W^\nu_{\rho\mu\sigma}\RE^\sigma_\nu \nonumber\\
&& +  \RE^\sigma_\mu\RE_{\sigma\rho} -{2\over3}\RE \RE_{\mu\rho} 
- \g_{\mu\rho}\left({\RE_{\sigma\rho}\RE^{\sigma\rho}\over4} - {\RE^2\over6}\right) 
\eeqa

The latter identities have a form well suited for an expansion of gravity theory around
GR. In the latter theory indeed, Einstein curvatures vanish everywhere
except on gravity sources, that is the Sun in the solar system.
This entails that identity (\ref{Bianchi_E}) thus has a simplified form 
\beqa
\label{Bianchi_deltaE}
\stand{\Partial_\lambda} \delta\RE^\lambda_{\mu} &\equiv&
\partial_\lambda \delta\RE^\lambda_{\mu} + 
\stand{\Gam^\lambda_{\lambda\rho}} \delta\RE^\rho_\mu -
\stand{\Gam^\rho_{\lambda\mu}} \delta\RE^\lambda_\rho=0 
\eeqa
The simplification is even more important for the identity (\ref{Bianchi_W}) 
with the disappearance of quadratic forms of Einstein curvatures
\beqa
\label{Bianchi_deltaW}
\delta\Partial_\lambda \W^\lambda_{\mu\nu\rho} &=&
{1\over2}\stand{\Partial_\nu}\delta\left(\RE_{\mu\rho} -{1\over3}\g_{\mu\rho}\RE\right)
-{1\over2}\stand{\Partial_\rho}\delta\left(\RE_{\mu\nu} -{1\over3}\g_{\mu\nu}\RE\right)\nonumber\\
\delta\Partial^\nu\Partial_\lambda \W^{\lambda}_{\mu\nu\rho} &=&
{1\over2}\stand{\Partial^2}\delta\RE_{\mu\rho} 
+{1\over6}\stand{\Partial_\mu\Partial_\rho-\g_{\mu\rho}\Partial^2}\delta\RE
- {1\over2}\stand{\W^\nu_{\rho\mu\sigma}}\delta\RE^\sigma_\nu \nonumber\\
\eeqa
Note that these equations are not valid on the gravitation sources.
They can be simplified one step further by using the fact
that potentials or curvatures have a relatively small amplitude 
in the solar system (see \ref{app:approx}).

\section{Approximations in the solar system}
\label{app:approx}

In this appendix, we write down approximated expressions by using two assumptions
which are valid in the solar system: the extended gravitation theory 
lies in the vicinity of GR and curvatures have a relatively small amplitude. 

The first assumption allows one to use expressions obtained in \ref{app:bianchi}. 
The second assumption then leads to a 
substitution of covariant derivatives by ordinary derivatives.
The transversality of Einstein curvatures (\ref{Bianchi_deltaE})
is thus simply read as the ordinary Minkowskian expression
\beqa
\label{Bianchi_deltaE_approx}
\partial_\lambda \delta\RE^\lambda_{\mu} =0 
\eeqa
Using also the other Bianchi identities (\ref{Bianchi_deltaW}), 
the transverse Einstein tensor may then be decomposed using projectors $\pros$ and $\prot$
\beqa
\label{sector_decomposition}
\delta\RE_{\mu\nu} &=& \delta\REs_{\mu\nu} + \delta\REt_{\mu\nu} \\
\delta\REs_{\mu\nu}&\equiv&\pros_{\mu\nu\lambda\rho}\delta\RE^{\lambda\rho} \quad , \quad
\delta\REt_{\mu\nu}\equiv\prot_{\mu\nu\lambda\rho}\delta\RE^{\lambda\rho} \nonumber\\
\pros_{\mu\nu\lambda\rho}&\equiv&
{\pro_{\mu\lambda}\pro_{\nu\rho} + \pro_{\lambda\nu}\pro_{\mu\rho}\over2}
-{\pro_{\mu\nu}\pro_{\lambda\rho}\over3}\nonumber \nonumber\\
\prot_{\mu\nu\lambda\rho} &\equiv&{\pro_{\mu\nu}\pro_{\lambda\rho}\over3} \quad , \quad
\pro_{\mu\nu} \equiv \eta_{\mu\nu} -{\partial_\mu\partial_\nu\over\partial^2}
\nonumber
\eeqa
Equations (\ref{Bianchi_deltaW}) relating Einstein curvatures to Weyl and scalar 
curvatutres then take a simple form
\beqa
\delta \RE_{\mu\nu}(\x) &=& -\pros_{\mu\nu00}\Delta \delta\wz(x)
- \prot_{\mu\nu00}\delta\R(\x)\nonumber\\
&&\Delta \equiv \partial_\r^2 + {2\over\r}\partial_\r
\eeqa
This reproduces the decomposition on the sectors of traceless and traced 
tensors which was used in \cite{JR05mpl,JR05cqg}.
We recall that the indices refer to the conformal weight of the components,
0 for traceless tensors and 1 for traced ones \cite{Jaekel95}.

The modification of gravitation equations may also be characterized in the momentum 
domain by introducing scale dependent running coupling gravitation constants $\Gts$ 
and $\Gtt$ which differ in the two sectors \cite{JR05mpl,JR05cqg}.
The perturbed Einstein gravitation equations (\ref{modified_Einstein_equations})
may indeed be rewritten (compare with sect.2 in \cite{JR05cqg}) 
\beqa
\label{modified_couplings}
\f(\x) &\equiv& \int {d^4\k \over (2\pi)^4}e^{-i\k\x} \f[\k] \quad, \quad 
\pro_{\mu\nu} \equiv \eta_{\mu\nu} -{\k_\mu\k_\nu\over\k^2}\nonumber\\
\delta \RE_{\mu\nu}[\k] &=& \pros_{\mu\nu00}{8\pi\delta\Gts[\k]\M\over\c^2}
+ \prot_{\mu\nu00}{8\pi\delta\Gtt[\k]\M\over\c^2} 
\eeqa
The information contained in the anomalies $\delta\Gts$ and $\delta\Gtt$ of the running
gravitation constants is completely equivalent to the spatial dependence of the curvature 
components, for instance $\delta \wz$ and $\delta\R$,
\beqa
\label{modified_couplings_R}
\Delta \delta\wz(x) &\simeq& -{8\pi\delta\Gts(x)\M\over\c^2} \, , \qquad
\delta\R(x) \simeq -{8\pi\delta\Gtt(x)\M\over\c^2}
\eeqa
The same information may equivalently be presented in terms of the modifications
of Einstein components 
\beqa
\label{modified_couplings_E}
\delta \RE^0_0 &\simeq& -{2\over3}\Delta\delta\wz -{1\over3}\delta\R
\simeq {2\delta\Gts + \delta\Gtt\over3}{8\pi\M\over\c^2}\nonumber\\ 
\delta \RE^\r_\r &\simeq& {2\over\r}\partial_\r{1\over3}{1\over\Delta}
(\Delta\delta\wz -\delta\R) \simeq{2\over\r}\partial_\r{1\over\Delta}
{\delta\Gtt - \delta\Gts\over3}{8\pi\M\over\c^2}
\eeqa
or, alternatively, in terms of two potentials $\PbN$ and $\PbP$
(see eqs.(\ref{Schwartzschild_metric_solution}); Schwartzschild coordinates are used)
\beqa
\label{linear_equivalence}
\Deltb \PbN &\simeq& {4\delta \Gts - \delta\Gtt\over 3}{4\pi\M\over\c^2}\quad, \quad 
\Deltb \equiv \partial_\lr^2 +{2\over\lr}\partial_\lr\nonumber\\
\Deltb \PbP &\simeq& {2(\delta \Gts - \delta\Gtt)\over 3}{4\pi\M\over\c^2}
\eeqa

In particular, potentials varying linearly with the radius \cite{JR05mpl,JR05cqg} 
or, equivalently, Einstein curvatures varying as the inverse of the radius correspond
to anomalous couplings $\delta\Gts[\k]$ and $\delta\Gtt[\k]$ scaling as $1/{\bf k}^2$ where
${\bf k}$ is the spatial wavevector.

\section{The PPN Ansatz}
\label{app:ppn}

In this appendix, we write down the PPN ansatz as a particular case of
the more general framework presented in the rest of the paper. 

The PPN metric is written in terms of isotropic coordinates 
(\ref{Eddington_isotropic_metric}) with
\beqa
\label{Eddington_PPN}
\g_{00} &=&1 + 2\pst + 2\bE \pst^2 +\ldots, 
\qquad \pst \equiv -\Gauss \u \nonumber\\
\g_{\r\r} &=& -1 + 2\cE\pst -{3\over2} \dE\pst^2 +\ldots
\eeqa
The Eddington PPN parameters $\bE$, $\cE$ and $\dE$ are equal to $1$ in general relativity
and close to $1$ in its vicinity. The symbol $\ldots$ has been introduced to recall that
these expressions are but the first terms in an expansion in terms of the Newton gravitation 
constant $\GN$ or, equivalently, of the Newton potential $\pst$. 

It is instructive to write down the expansion of the anomalous Einstein curvatures 
\beqa
\label{PPN_curvatures}
\delta\RE^0_0\simeq&& 3\left(2(\cE-1) -(\dE-1)\right)\Gauss ^2 \u^4 
+\ldots \nonumber\\
\delta\RE^\r_\r\simeq&& 2 (\cE-1) \Gauss  \u^3
+\left( 4(\bE-1)-12(\cE-1)+3(\dE-1)\right) \Gauss ^2 \u^4+\ldots 
\eeqa
This expansion shows that $\delta\RE^0_0$ vanishes at leading order in $\GN$,
whereas $\delta\RE^\r_\r$ does not.
In other words, the PPN metric is a particular extension of GR
which affects $\delta\RE^\r_\r$ more than $\delta\RE^0_0$.

It is also worth writing the anomalous parts of the two potentials $\PbN$ and $\PbP$ 
for the PPN metric
\beqa
\label{PPN_potentials}
&&\PbN^{(PPN)} \simeq (\bE-1)\Gauss^2\ub^2\\
&&\PbN^{(PPN)} - \PbP^{(PPN)} \simeq - (\cE-1)\Gauss\ub +\left( {3\over2}(\cE-1) - {3\over4}(\dE-1)\right)
\Gauss^2\ub^2\nonumber
\eeqa
This shows that $\PbN$ and $\PbP$ vary quadratically with the inverse radius.
In other words, the PPN metric is a particular extension of GR
with anomalies present at short ranges (in the inner solar system) rather 
than at long ranges (in the outer solar system).

\section{Precession anomaly}
\label{App:precession}

In GR, the gravitation potential takes its standard Newtonian form. We may then solve
the Kepler problem by introducing a polynomial $\stand{\V}$ of third degree in the 
inverse radial coordinate (Schwartzschild coordinates are used in this appendix), 
with coefficients determined by the two invariants $\J$ and $\E$ 
\beqa
&&\stand{\V}(\ub)\equiv {\E^2\over\J^2\c^2} 
- \stand{\gb_{00}}\left({\m^2\c^2\over\J^2} +\ub^2\right)
\eeqa
We may also write it in terms of the Keplerian elements, namely the orbit mean inverse 
radius $\ux$ and the orbit eccentricity $\eb$,
\beqa
\label{Einstein_potential}
&&\stand{\V}(\ub)
=\Om^2\left( \eb^2\ux^2 -(\ub-\ux)^2 + {2\db\over\ux}(\ub-\ux)^3\right)
\nonumber\\
&&\ux\left(1+{3\over2}\stand{\gb_{00}}^\prime\ux\right) \equiv -{1\over2}\stand{\gb_{00}}^\prime{\m^2\c^2\over\J^2}\nonumber\\
&&\Om^2 = {1\over 1+6\db} \equiv 1 +3\stand{\gb_{00}}^\prime \ux \nonumber\\
&&\Om^2\left(\eb^2 - 1 - 2\db\right)\ux^2 \equiv {\E^2\over\J^2\c^2} 
- {\m^2\c^2\over\J^2} 
\eeqa
Hamilton-Jacobi equations (\ref{Hamilton_Jacobi}) lead to the following
solution 
\beqa
\label{Einstein_geodesic}
\stand{\Sr} &=& \mp \J \int^{\ub} {\sqrt{\stand{\V}}\over \stand{\gb_{00}}} {d\ub \over\ub^2}
\quad,\quad
\stand{\ang} = \pm \int^\ub {d\ub \over\sqrt{\stand{\V}}}
\eeqa
which allows one to recover the perihelion precession for Einstein geodesics
\beqa
\label{precessions}
2\pi + \stand{\Dang} &=& \intc d\stand{\ang} = \pm \intc  {d\ub \over\sqrt{\stand{\V}}} \\
&=&\pm {1\over \Om}\int_0^{2\pi} {(1-2\eb\db{\rm{sin}}\ang)^{-{1\over2}} {\rm{cos}}\ang 
d\ang\over\sqrt{{\rm{cos}}^2\ang +{1\over\eb^2\db^2}(1-\eb\db-
\sqrt{1-2\eb\db{\rm{sin}}\ang})^2}} \nonumber
\eeqa

We come now to the post-Einsteinian case, and expand 
the generating function for precessions, namely $\DSr$ (\ref{action_precession}), 
in the vicinity of GR metric (\ref{perturbed_Einstein_geodesic}). 
This function depends on the constants of motion 
and metric components through two generic integrals
\beqa
\label{perturbed_precessions}
\delta \DSr &\equiv& -{\J\over2}\Isst{\stand{\gb_{00}}\delta\gb_{\r\r}}
+{\E^2\over2\J\c^2}\Ipst{{\delta\gb_{00} \over\ub^2\stand{\gb_{00}}^2}} \\
\Isst{\f} &\equiv& \intc \f(\ub) {d\stand{\Sr}\over\J}\quad,\quad
\Ipst{\f} \equiv \intc \f(\ub) d\stand{\ang} \nonumber
\eeqa
Corrections brought by the Newtonian potential are of order 
$|\stand{\gb_{00}}^\prime\ux| \ll 1$
and may be calculated in a perturbative expansion. 
Integrals in equations (\ref{perturbed_precessions}) are thus expanded around 
expressions for Keplerian orbits ($\Om = 1, \db = 0$). Noting that 
$\E^2/\J^2\c^2\ux^2$ is of order $-1$ (see (\ref{Einstein_potential})),
 one rewrites expressions in (\ref{perturbed_precessions}) in terms of 
$\ux$ and $\eb$
\beqa
\label{perturbed_precession_approximation}
&&{\E^2\over\J^2\c^2} \simeq \ux^2\left( -{2\over\stand{\gb_{00}}^\prime\ux} 
+ \eb^2 -4\right)\quad, \quad \db \simeq -{1\over2}\stand{\gb_{00}}^\prime\ux\nonumber\\
&&\ub\simeq \ux\left(1+\eb{\rm{sin}}\ang +{\db\over2}\eb^2{\rm{\sin}}^2\ang\right)\quad,\quad
d\stand{\ang} \simeq \left( 1+\db(3+\eb{\rm{sin}}\ang)\right)
d\ang\nonumber\\
&&\I{\f} \equiv \int_0^{2\pi}\f\left(\ux(1+\eb{\rm{\sin}}\ang)\right)d\ang
\quad,\quad
\Isst{\f}\simeq  \I{[(\ub-\ux)^2 - \eb^2\ux^2]{\f\over\ub^2}}\nonumber\\
&&\Ipst{\f}\simeq (1+3\db)\I{\f} + {\db\over2\ux}\I{((\ub-\ux)^2\f)^\prime}
\eeqa

In fact, confrontation with observations shows that the two potential perturbations 
do not occur at the same level, but that corrections to the first potential occur
at one order higher than to the second potential. Taking this remark into account, 
and using relations (\ref{Einstein_potential}), we obtain
from (\ref{perturbed_precessions}) and (\ref{perturbed_precession_approximation}) 
the precession anomaly at leading order 
\beqa
\label{elements_derivatives}
&&\delta \Dang \simeq -{1\over2}(1+\J\partial_\J)\Isst{\delta\gb_{\r\r}}
+ {\ux\over\stand{\gb_{00}}^\prime}
(1-\J\partial_\J)\I{{\delta\gb_{00}\over\ub^2}}
\nonumber\\
&&\J\partial_\J \simeq -2\left(\ux\partial_{\ux} +(1-\eb^2)\partial_{\eb^2}\right)
\eeqa

For a nearly circular orbit, expressions (\ref{perturbed_precession_approximation})
may be expanded in the orbit eccentricity $\eb$
\beqa
\label{Kepler_secular_integrals}
&&-{1\over2}(1+\J\partial_\J)\Isst{\f} \simeq -\pi \left( \f(\ux) + {\eb^2\ux^2\over4}
\f^{\prime\prime}(\ux)\right)\nonumber\\
&&(1-\J\partial_\J)\I{\f} \simeq 
\pi\left((\ub^2\f)^{\prime\prime}(\ux)
+{\eb^2\ux^2\over8}(\ub^2\f)^{(4)}(\ux)\right)
\eeqa
One then obtains the perihelion precession anomaly at leading order
\beqa
\label{precession_anomaly}
\delta \Dang &\simeq& -\pi\left( \delta\gb_{\r\r}
-{\ux\delta\gb_{00}^{\prime\prime}\over\stand{\gb_{00}}^\prime}
+{\eb^2\ux^2\over4}\left(\delta\gb_{\r\r}^{\prime\prime}
-{\ux\delta\gb_{00}^{(4)}\over2\stand{\gb_{00}}^\prime}\right)\right)
\eeqa
Functions of the radial coordinate have been evaluated at $1/\ux$.

\def\etal{\textit{et al }}
\def\ibid{\textit{ibidem }}
\def\eprint#1{{\it Preprint} #1}

\end{document}